\documentclass
[aps,pre,showpacs,showkeys,floatfix,nofootinbib,longbibliography,onecolumn]{revtex4-1}%
\usepackage{amsfonts}
\usepackage{amsmath}
\usepackage{amssymb}
\usepackage{graphicx}
\usepackage{verbatim}
\usepackage{subfigure}
\usepackage{bm}%
\providecommand{\U}[1]{\protect\rule{.1in}{.1in}}
\newcommand{\ecc}{\epsilon}
\newcommand{\Vea}{V_{exc}^{(\mathrm{app})}}
\DeclareMathOperator\arctanh{arctanh}

\begin{document}
\title{Density functional theory for dense nematics with steric interactions}
\author{Eduardo S. \surname{Nascimento}}
\email{[e-mail: ]enascime@kent.edu}
\affiliation{Liquid Crystal Institute, Kent State University, OH, USA}
\author{Peter \surname{Palffy-Muhoray}}
\email{[e-mail: ]mpalffy@kent.edu}
\affiliation{Liquid Crystal Institute, Kent State University, OH, USA}
\author{Jamie M. \surname{Taylor}}
\email{[e-mail: ]Jamie.Taylor@maths.ox.ac.uk}
\affiliation{Mathematical Institute, University of Oxford, Oxford, UK }
\author{Epifanio G. \surname{Virga}}
\email{[e-mail: ]eg.virga@unipv.it}
\thanks{On leave from Dipartimento di Matematica, Universit\`a di Pavia, Via Ferrata
5, I-27100 Pavia, Italy}
\affiliation{Mathematical Institute, University of Oxford, Oxford, UK }
\author{Xiaoyu \surname{Zheng}}
\email{[e-mail: ]zheng@math.kent.edu}
\affiliation{Department of Mathematical Sciences, Kent State University, OH, USA}

\begin{abstract}
The celebrated work of Onsager on hard particle systems,
based on the truncated second order virial expansion, is valid at relatively
low volume fractions for large aspect ratio particles. While it predicts the
isotropic-nematic phase transition, it does not provide a realistic equation
of state in that the pressure remains finite for arbitrarily high densities.
In this work, we derive a mean field density functional form of the Helmholtz
free energy for nematics with hard core repulsion. In addition to predicting
the isotropic-nematic transition, the model provides a more realistic equation
of state. The energy landscape is much richer, and the orientational
probability distribution function in the nematic phase possesses a unique
feature---it vanishes on a nonzero measure set in orientation space.

\end{abstract}
\maketitle

\section{Introduction}

\label{sec:intro} An ensemble of non-spherical particles, interacting via hard
core interactions, exhibits a first order isotropic-nematic phase transition
as the number density increases, followed by a nematic-solid transition
\cite{frenkel:phase,Frenkel85} (see also the companion paper
\cite{mulder:hard}, for a number of technical results on the excluded volume
of ellipsoids). In his celebrated theoretical work \cite{Onsager49}, Onsager
successfully predicted the isotropic-nematic transition. Using the canonical
ensemble, he expressed the free energy in terms of the orientational
probability density function $f(\mathbf{\hat{l}})$, where $\mathbf{\hat{l}}%
\in\mathbb{S}^{2}$ is a unit vector along the symmetry axis of a particle.
Truncating the virial expansion after the second term, in the low volume
fraction limit, the free energy is of the form
\begin{equation}
\label{eq_onsager}F=NkT\left\{  \ln\rho_{0}+\int_{\mathbb{S}^{2}%
}f(\mathbf{\hat{l}})\left(  \ln f(\mathbf{\hat{l}})+\frac{1}{2}\rho_{0}%
\int_{\mathbb{S}^{2}}f(\mathbf{\hat{l}}^{\prime})V_{ex}(\mathbf{\hat{l}%
},\mathbf{\hat{l}}^{\prime})d\mathbf{\hat{l}}^{\prime}\right)  d\mathbf{\hat
{l}}\right\}  ,
\end{equation}
where $N$ is the number of particles, $k$ is Boltzmann's constant, $T$ is the
temperature and $\rho_{0}$ is the number density.\footnote{A justification of
Eq.~(\ref{eq_onsager}) can also be given in terms of a reduced cluster
expansion, based only on Penrose's spanning trees, which explains the success
of Onsager's theory better than the omission of all terms in the virial
expansion but the first \cite{Virga17}.} In addition to the isotropic-nematic
transition, the model describes phase separation and the coexistence of
nematic and isotropic phases. In spite of its many successes
\cite{mederos:hard}, the theory has a major limitation: it does not provide a
reasonable description of the behavior near the dense packing limit. It does
not give a reasonable equation of state; the pressure remains finite at
arbitrarily high number densities.

Onsager's theory was rooted in the Mayer expansion, which formed the basis of
Mayer's statistical mechanics book \cite{mayer:statistical}. Seen in the light
of the full Mayer expansion (and the related problem of its convergence) the
truncation at the first power in the number density in \eqref{eq_onsager}
appears unjustified. Many attempts have been made to explore the limits of
validity of the Onsager theory and to propose extensions.

Straley~\cite{straley:gas,straley:third}, by combining analytical and
numerical methods, estimated the third virial coefficient for hard rods and
concluded that Onsager's theory would not be quantitatively accurate for
values of the length-to-breadth ratio $\kappa$ less than $100$, a prediction
that was far too pessimistic and that was indeed proven to be unrealistic by
the Monte Carlo simulations of Frenkel and
Mulder~\cite{frenkel:phase,Frenkel85} with ellipsoids of revolution. They
showed that an ordering transition does take place already for ellipsoids of
aspect ratio $\kappa=3$, though the densities involved could be too high to be
compatible with the Onsager theory. It then became clear that it is not the
degree of anisometry of the particles, but their density that may more easily
challenge the validity of Onsager's theory. The emphasis thus shifted from the
particles' shape to their filling fraction. A similar conclusion was drawn
much later in the work of Tjipto-Margo and Evans~\cite{tjipto-margo:onsager},
who included the third virial coefficient in the theory for ellipsoids. For
aspect ratios $\kappa>5$, their extended theory predicts the correct
variation of the order parameter with density and is in agreement with simulations.

A valuable improvement to the Mayer paradigm of density expansion was
introduced by Barboy and Gilbert~\cite{barboy:series}. They remarked that an
expansion in the variable $y = \eta/ (1-\eta)$, where $\eta$ denotes the
volume fraction, has better convergence properties than the usual series in
the number density. Their theory was applied to dumbbells and spherocylinders
in the isotropic phase~\cite{barboy:series}, as well as to
hard-parallelepipeds~\cite{barboy:hard}, but in the restricted-orientation
approximation of Zwanzig~\cite{zwanzig:first}. No clear superiority of this
theory to the classical Mayer theory was ever established with
certainty~\cite{mederos:hard}.

A considerable step towards the extension of Mayer theory was taken by
Parsons~\cite{parsons:nematic} and Lee~\cite{lee:numerical}, who moved along
different lines of thought but arrived at one and the same theory. In Lee's
interpretation, the emphasis is on the equation of state, which is modelled
after the theory of Carnahan and Starling~\cite{carnahan:equation} for real
gases of hard spheres. The virial coefficients for the pressure expansion of
Mayer's theory are rescaled to those hypothesised by Carnahan and Starling in
their extrapolation of the viral coefficients for hard spheres known at that
time. As a consequence of this rescaling, the expansion for the free energy
functional is re-summed exactly and expressed in the form of a modified
Onsager bilinear functional in the probability distribution density. Lee
applied this theory to a fluid of axially symmetric ellipsoids and found it in
a better agreement with simulations than other theories available at that
time~\cite{lee:onsager}. An alternative theory was independently proposed by
Baus and co-workers~\cite{baus:finite,colot:density}, which despite
appearances is easily seen to reduce to the Parsons-Lee theory
\cite{mederos:hard}.

As successful as the Parsons-Lee theory may be, it is limited by being a
re-summed version of the Mayer theory, and in so it remains exposed to all the
unresolved issues concerned with the convergence of the underlying Mayer power
series. We therefore attempt here to go beyond the Mayer paradigm.

In this paper, the density functional theory of aspherical hard particles is
revisited. We are particularly interested in the phase behavior of hard
ellipsoids at high densities, and near the densest packing limit. This very
system will be used as a test case for the theory. Within the comfort zone of
Mayer expansion, the evaluation of the partition function at high densities,
even for hard spheres, has eluded researchers to date in spite of considerable
effort. Here we take the simplest approach to explore orientational order in
this regime. To clearly focus on this problem, we have chosen a minimal model,
essentially at the van der Waals level. Although one might suspect that this
form cannot provide quantitative predictions, our comparison with simulations
for ellipsoids is encouraging. More generally, we feel that, in spite of its
simplicity, it can give valuable insights into the consequences of the salient
aspect of the problem: the depletion of available phase space as the density
is increased.

A major challenge is the determination of the orientational distribution
function subject to the hard constraint that the number of available states to
the system is positive. This constraint is not always strictly enforced in the
literature~\cite{PPM}. One striking result of this constraint, in the mean
field limit, is that at densities above the isotropic-nematic transition, the
orientational distribution function which minimizes the free energy has a
compact support over orientation space. We also find that the nematic phase is
more orientationally ordered than in the Onsager theory at the same density,
and the density of the coexisting isotropic and nematic phases is a simple
function of the particle shape. Our simple model gives a reasonable equation
of state; that is, the pressure diverges at the dense packing limit.

The goal of the paper is to provide an approximate but near realistic
description of dense orientationally ordered hard particle systems; currently
available descriptions are either low density approximations \cite{Sheng75,
Herzfeld}, or they rely on somewhat \textit{ad hoc} Carnahan-Starling type
corrections to the free energy at high densities \cite{Jak1, Jak2}.

In Sec.~\ref{sec:theory}, we derive our theory. In Sec.~\ref{sec:ellipsoids},
we write the free energy for a system of hard ellipsoids with arbitrary number
density, derive and solve the Euler-Lagrange equation for the probability
density function, and arrive at the equation of state. In
Sec.~\ref{sec:assumption_uniaxiality}, we assume uniaxial nematic order, and
present some numerical results relevant to this case. In
Sec.~\ref{phase_coexistence}, we discuss phase separation and two-phase
coexistence. In Sec.~\ref{sec:biaxial}, we explore the possibility of biaxial
equilibrium phases. Section~\ref{sec:comparison} is devoted to a close
comparison with simulations. Finally, in Sec.~\ref{sec:discussion} we draw the
conclusions of this study and summarize our results. The paper is closed
by an Appendix, where we justify the approximate expression used in
Sec.~\ref{sec:ellipsoids} for the excluded volume of ellipsoids of revolution.

\section{Theory}

\label{sec:theory} We consider a one-component system consisting of hard
particles. For simplicity, we omit attractive interactions. The
configurational Helmholtz free energy of a system of $N$ particles, within an
additive constant, is
\begin{equation}
F=-kT\ln\frac{1}{N!}\int_{\Omega^{N}}e^{-\frac{1}{kT}\sum_{1\leq i<j\leq
N}U_{ij}^{R}}d\mathbf{q}_{1}\cdots d\mathbf{q}_{N},
\end{equation}
where $\mathbf{q}_{i}$ is the generalized (orientational and positional)
coordinate of the $i^{th}$ particle, and $U_{ij}^{R}=U^{R}(\mathbf{q}_{i},
\mathbf{q}_{j})$ is the repulsive interaction energy between particles $i$ and
$j$, the sum is over all pairs of particles. Explicitly, the interaction
potential is
\begin{equation}
U^{R}(\mathbf{q}_{i},\mathbf{q}_{j})=
\begin{cases}
\infty, & \text{if particles interpenetrate},\\
\  & \\
0, & \text{otherwise}.
\end{cases}
\end{equation}

We write
\begin{equation}
G_{N}=\int_{\Omega^{N}}e^{-\frac{1}{kT}\Sigma_{1\leq i<j\leq N}U_{ij}^{R}%
}d\mathbf{q}_{1}\cdots d\mathbf{q}_{N},
\end{equation}
where the quantity $G_{N}$ can be thought of as the number of states available
to $N$ distinguishable particles. We consider adding one particle to the
system. Then
\begin{equation}
G_{N+1}=\int_{\Omega^{N}}e^{-\frac{1}{kT}\Sigma_{1\leq i<j\leq N}U_{ij}^{R}%
}\left(  \int_{\Omega}e^{-\frac{1}{kT}\Sigma_{i=1}^{N}U_{i(N+1)}^{R}%
}d\mathbf{q}_{N+1}\right)  d\mathbf{q}_{1}\cdots d\mathbf{q}_{N},
\end{equation}
and, since the probability density function of a given configuration is
\begin{equation}
P(\mathbf{q}_{1},\cdots,\mathbf{q}_{N})=\frac{e^{-\frac{1}{kT}\Sigma_{1\leq
i<j\leq N}U_{ij}^{R}}}{G_{N}},
\end{equation}
we have
\begin{align}
G_{N+1}  &  =G_{N}\int_{\Omega^{N}}P(\mathbf{q}_{1},\cdots,\mathbf{q}%
_{N})\left(  \int_{\Omega}e^{-\frac{1}{kT}\Sigma_{i=1}^{N}U_{i(N+1)}^{R}%
}d\mathbf{q}_{N+1}\right)  d\mathbf{q}_{1}\cdots d\mathbf{q}_{N}\nonumber\\
&  =G_{N}\left\langle \int_{\Omega}e^{-\frac{1}{kT}\Sigma_{i=1}^{N}%
U_{i(N+1)}^{R}}d\mathbf{q}_{N+1}\right\rangle ,
\end{align}
where the average is computed relative to $P$. Since all particles are
equivalent, we may write
\begin{equation}
G_{N}=\left\langle \int_{\Omega}e^{-\frac{1}{kT}\Sigma_{i=2}^{N}U_{1i}^{R}%
}d\mathbf{q}_{1}\right\rangle ^{N},
\end{equation}
where the integral represents the average free volume per particle in an
ensemble of $N$ particles. We can equivalently write this as
\begin{equation}%
\begin{split}
G_{N}  &  =\left\langle \int_{\Omega}1-(1-e^{-\frac{1}{kT}\Sigma_{i=2}%
^{N}U_{1i}^{R}})d\mathbf{q}_{1}\right\rangle ^{N}\\
&  =\left(  \int_{\Omega}[1-W(\mathbf{q}_{1})]d\mathbf{q}_{1}\right)  ^{N},
\end{split}
\end{equation}
where
\begin{equation}
W(\mathbf{q}_{1})=\left\langle 1-e^{-\frac{1}{kT}\Sigma_{i=2}^{N}U_{1i}^{R}%
}\right\rangle \label{w1}%
\end{equation}
is the average excluded volume fraction.

The free energy then becomes
\begin{equation}
F=-kT\ln\frac{1}{N!}\left(  \int_{\Omega}[1-W(\mathbf{q}_{1})]d\mathbf{q}%
_{1}\right)  ^{N}. \label{2}%
\end{equation}
This important result is exact. The integrand in Eq.~(\ref{2}) can be regarded
as the fraction of the total volume available to particle $1$, or the average
free volume fraction. We next write that
\begin{equation}
W(\mathbf{q}_{1})=\frac{N}{V}v_{eff},%
\end{equation}
where $v_{eff}$ is the average volume effectively occupied by one particle.
For two hard spheres of volume $v_{0}$, the pair excluded volume $V_{ex}$ is
\begin{equation}
V_{ex}=8v_{0},%
\end{equation}
and for $N=2$, Eq.~(\ref{w1}) gives exactly
\begin{equation}
v_{eff}=\frac{1}{2}V_{ex}.%
\end{equation}
This is likely to remain a good approximation in the low density limit, as
used in the Van der Waals equation \cite[pp.~90--91]{Israelachvili}, but it is
clear from Eq.~(\ref{w1}) as well as from experiments, that at high densities
$v_{eff}$ is reduced considerably from this value. For close packed spheres,
for example,
\begin{equation}
v_{eff}=\frac{3}{4\pi\sqrt{2}}V_{ex}\simeq\frac{1}{6}V_{ex}.
\label{eq:close_packed_spheres}%
\end{equation}
We write therefore that
\begin{equation}
v_{eff}=\lambda V_{ex},
\end{equation}
where $\lambda$ is a parameter, taken to be nearly constant in our high
density limit, to be determined by comparison with experiment.\footnote{This
parameter $\lambda$ is not dissimilar from the \emph{coordination} parameter
$z$ introduced in the generalization presented in \cite{gartland:minimum} of
the classical mean field theory as incarnated in \cite{palffy:single} (see
also Chapter 1 of \cite{sonnet:dissipative} for a more detailed discussion).}
This justifies the following postulate of our theory:
\begin{equation}
W(\mathbf{q}_{1})=\lambda\int_{\Omega}\rho(\mathbf{q}_{2})\left(
1-e^{-\frac{U^{R}(\mathbf{q}_{1},\mathbf{q}_{2})}{kT}}\right)  d\mathbf{q}%
_{2}, \label{eq:lambda}%
\end{equation}
where $\rho(\mathbf{q})$ is the number density of particles with generalized
coordinate $\mathbf{q}$. More generally, in Eq.~\eqref{eq:lambda} $\lambda$
should be thought of as a function of the number density $\rho(\mathbf{q}%
_{1})$, for whose form we have yet to present a firm first-principles derivation. Here,
$\lambda$ will be treated as an adjustable parameter; comparison with
simulations will indicate the range within which the function $\lambda(\rho)$
is expected to vary.

To obtain the density functional form of the free energy, we assume that the
density $\rho(\mathbf{q}_{i})$ is a slowly varying function of $\mathbf{q}%
_{i}$. \ We consider an element $\Omega_{i}$ of phase space, containing
$N_{i}$ particles and having volume $\Delta\mathbf{q}_{i}$, sufficiently small
so that $\rho(\mathbf{q}_{i})$ in element $\Omega_{i}$ is nearly constant. The
free energy $F_{i}$ of region $\Omega_{i}$ is of the form of Eq.~(\ref{2});
that is, by Stirling's approximation,\footnote{To be precise, the term
$-kTN_{i}$ has been omitted on the right-hand side of Eq.~(\ref{eq_14}). Its
inclusion would only add a term proportional to the total number of particles
to the right hand side of Eq.~(\ref{eq_17}).}
\begin{equation}
F_{i}=-kT\ln\left(  \frac{1}{N_{i}}\int_{\Omega_{i}}[1-W(\mathbf{q}%
_{i})]d\mathbf{q}_{i}\right)  ^{N_{i}}, \label{eq_14}%
\end{equation}
and since $\rho(\mathbf{q}_{i})$ is nearly constant, we have that $N_{i}%
=\rho(\mathbf{q}_{i})\Delta\mathbf{q}_{i}$, and
\begin{equation}
F_{i}=-kT\ln\left(  \frac{1}{\rho(\mathbf{q}_{i})}[1-W(\mathbf{q}%
_{i})]\right)  ^{\rho(\mathbf{q}_{i})\Delta\mathbf{q}_{i}}.
\end{equation}
Then, since the free energy is additive, we can write for the entire system
\begin{equation}
F=-kT\ln{\prod}_{i}\left(  \frac{1}{\rho(\mathbf{q}_{i})}[1-W(\mathbf{q}%
_{i})]\right)  ^{\rho(\mathbf{q}_{i})\Delta\mathbf{q}_{i}},
\end{equation}
or, passing to the continuum limit,
\begin{equation}
F=kT\int_{\Omega}\rho(\mathbf{q})\ln\rho(\mathbf{q})d\mathbf{q}-kT\int%
_{\Omega}\rho(\mathbf{q})\ln[1-W(\mathbf{q})]d\mathbf{q}, \label{eq_17}%
\end{equation}
which is the standard density functional form. Explicitly, this is
\begin{equation}
F=kT\int_{\Omega}\rho(\mathbf{q}_{1})\ln\rho(\mathbf{q}_{1})d\mathbf{q}%
_{1}-kT\int_{\Omega}\rho(\mathbf{q}_{1})\ln\left[  1-\lambda\int_{\Omega}%
\rho(\mathbf{q}_{2})\left(  1-e^{-\frac{U^{R}(\mathbf{q}_{1},\mathbf{q}_{2}%
)}{kT}}\right)  d\mathbf{q}_{2}\right]  d\mathbf{q}_{1}.
\label{eq:pre_free_energy}%
\end{equation}

For our problem, it is convenient to write the generalized coordinates in
terms of position and orientation, then
\begin{equation}%
\begin{split}
F  &  =kT\int_{\mathbb{S}^{2}}\int_{\mathcal{B}}\rho(\mathbf{r}_{1}%
,\mathbf{\hat{l}}_{1})\ln\rho(\mathbf{r}_{1},\mathbf{\hat{l}}_{1}%
)d^{3}\mathbf{r}_{1}d^{2}\mathbf{\hat{l}}_{1}\\
&  -kT\int_{\mathbb{S}^{2}}\int_{\mathcal{B}}\rho(\mathbf{r}_{1}%
,\mathbf{\hat{l}}_{1})\ln\left[  1-\lambda\int_{\mathbb{S}^{2}}\int%
_{\mathcal{B}}\rho(\mathbf{r}_{2},\mathbf{\hat{l}}_{2})\left(  1-e^{-\frac
{U^{R}(\mathbf{r}_{1},\mathbf{\hat{l}}_{1},\mathbf{r}_{2},\mathbf{\hat{l}}%
_{2})}{kT}}\right)  d^{3}\mathbf{r}_{2}d^{2}\mathbf{\hat{l}}_{2}\right]
d^{3}\mathbf{r}_{1}d^{2}\mathbf{\hat{l}}_{1},
\end{split}
\label{eq:free_energy}%
\end{equation}
where $\rho(\mathbf{r},\mathbf{\hat{l}})$ is the number density of particles
with center of mass at position $\mathbf{r}$, and orientation of symmetry axis
along $\mathbf{\hat{l}}$. The unit sphere $\mathbb{S}^{2}$ is the orientation
space. The region $\mathcal{B}$ in physical space denotes the position space,
occupied by particles. In a homogeneous system, the density is independent of
$\mathbf{r}$, so we write
\begin{equation}
\rho(\mathbf{r},\mathbf{\hat{l}})=\rho_{0}f(\mathbf{\hat{l}}),
\end{equation}
where $\rho_{0}$ is now simply the number density of particles, and
$f(\mathbf{\hat{l}})$ is the single particle orientational distribution
function satisfying
\begin{equation}
\int_{\mathbb{S}^{2}}f(\mathbf{\hat{l}})d^{2}\mathbf{\hat{l}}=1.
\end{equation}
Integrating over $\mathbf{r}_{2}$ in Eq.~\eqref{eq:free_energy} gives
\begin{equation}
\int_{\mathbb{S}^{2}}\int_{\mathcal{B}}\rho_{0}f(\mathbf{\hat{l}}_{2})\left(
1-e^{-\frac{U^{R}(\mathbf{r}_{1},\mathbf{\hat{l}}_{1},\mathbf{r}%
_{2},\mathbf{\hat{l}}_{2})}{kT}}\right)  d^{3}\mathbf{r}_{2}d^{2}%
\mathbf{\hat{l}}_{2}=\rho_{0}\int_{\mathbb{S}^{2}}f(\mathbf{\hat{l}}%
_{2})V_{exc}(\mathbf{\hat{l}}_{1},\mathbf{\hat{l}}_{2})d^{2}\mathbf{\hat{l}%
}_{2},
\end{equation}
where
\begin{equation}
V_{exc}(\mathbf{\hat{l}}_{1},\mathbf{\hat{l}}_{2})=\int_{\mathcal{B}}\left(
1-e^{-\frac{U^{R}(\mathbf{r}_{1},\mathbf{\hat{l}}_{1},\mathbf{r}%
_{2},\mathbf{\hat{l}}_{2})}{kT}}\right)  d^{3}\mathbf{r},
\end{equation}
is the excluded volume of two particles, which depends only on the relative
orientation of the particles with respect to one another. The free energy in
Eq.~\eqref{eq:pre_free_energy} then becomes
\begin{equation}
F=kT\rho_{0}V\left[  \ln\rho_{0}+\int_{\mathbb{S}^{2}}f(\mathbf{\hat{l}}%
_{1})\ln f(\mathbf{\hat{l}}_{1})d^{2}\mathbf{\hat{l}}_{1}-\int_{\mathbb{S}%
^{2}}f(\mathbf{\hat{l}}_{1})\ln\left(  1-\lambda\rho_{0}\int_{\mathbb{S}^{2}%
}f(\mathbf{\hat{l}}_{2})V_{exc}(\mathbf{\hat{l}}_{1},\mathbf{\hat{l}}%
_{2})d^{2}\mathbf{\hat{l}}_{2}\right)  d^{2}\mathbf{\hat{l}}_{1}\right]  .
\label{eq_energy}%
\end{equation}
If $\rho_{0}$ is small, one can expand the logarithm and recover the theory of
Onsager~\cite{Onsager49} as well as that of Doi and Edwards \cite[p.~354]%
{Doi&Edwards},\footnote{For a gas of hard spheres, van Kampen \cite{Kampen64}
had already proposed a free energy that features the logarithm of the free
volume. However, his derivation, which admittedly follows closely unpublished
work of L.S. Ornstein (1908), is \textit{ad hoc} and finds implicit
inspiration in the exact treatment of one dimensional Tonks' gas
\cite{Tonks36}.} We note that the expansion is only valid when
\begin{equation}
\lambda\rho_{0}\int_{\mathbb{S}^{2}}f(\mathbf{\hat{l}}_{2})V_{exc}%
(\mathbf{\hat{l}}_{1},\mathbf{\hat{l}}_{2})d^{2}\mathbf{\hat{l}}_{2}<1.
\end{equation}
In general, the argument of the logarithm must be positive. The equilibrium
orientational distribution $f(\mathbf{\hat{l}})$ is a function of $\rho_{0}$,
and is not known \textit{a priori}. Intuitively, one would expect particles to
align more and more as the number density is increased, corresponding to a
decrease of the orientationally averaged excluded volume with number density.
It follows that the densest packing density cannot be greater than the inverse
of the smallest average excluded volume of a pair of particles. The
orientational distribution function is therefore expected to depend
sensitively on the number density $\rho_{0}$.

To capture the phase behavior in the high density regime, we keep the full
logarithmic dependence in Eq.~\eqref{eq_energy}; this is the salient feature
of our approach. This results in a remarkable phenomenon: above a critical
value of $\rho_{0}$, the equilibrium distribution function $f(\mathbf{\hat{l}%
})$ is continuous over the whole orientation space $\mathbb{S}^{2}$, but
vanishes on a region with positive measure; that is, at high densities, some
regions of orientation space are not accessible to particles.

\section{Application to Ellipsoids}

\label{sec:ellipsoids} In this section we apply the theory presented in the
Sec.~\ref{sec:theory} to a system of hard ellipsoids of revolution. To this
end, our first task is to obtain a simple but reliable expression for the
excluded volume of two such particles.

\subsection{Excluded volume}

For identical hard ellipsoids of revolution of length $L$, width $W$, the
volume is $v_{0}=\frac{1}{6}\pi L^{3}/\kappa^{2}$, where $\kappa=L/W$ is the
\emph{aspect ratio}. A simple approximate expression for the excluded volume
is derived in the Appendix, where its accuracy is also assessed; it reads as
\begin{align}
V_{exc}(\mathbf{\hat{l}}_{1},\mathbf{\hat{l}}_{2})  &  =C-\frac{2}{3}D\left(
\mathbf{\sigma(\hat{l}}_{1}):\mathbf{\sigma(\hat{l}}_{2}) \right)
=C-DP_{2}(\mathbf{\hat{l}}_{1}\cdot\mathbf{\hat{l}}_{2})
,\label{eq_excluded_vol}\\
C  &  =8v_{0}(1+\varphi),\label{eq_c}\\
D  &  =8v_{0}\varphi, \label{eq_d}%
\end{align}
where $\mathbf{\bm{\sigma}(\hat{l})=}\frac{1}{2}(3\mathbf{\hat{l}\hat{l}%
}-\mathbf{I)}$ denotes the orientation descriptor of a particle with symmetry
axis oriented along $\mathbf{\hat{l}}$ and $P_{2}(x)=\frac32x^{2}-\frac12$ is
the second Legendre polynomial. The shape parameters $C\geq0$ and $C>D$ are
expressed in terms of a convenient measure of the anisotropy measured by
$\varphi$, which in Eq.~\eqref{eq:phi_function} is given as an explicit,
though complicated positive function of the eccentricity $\epsilon$ of the
ellipsoid, defined as
\begin{equation}
\label{eq:eccentricity}\epsilon:=
\begin{cases}
\sqrt{1-\kappa^{2}}, & \text{for}\ 0\leqq\kappa\leqq1,\\
\  & \\
\sqrt{1-\frac{1}{\kappa^{2}}}, & \text{for}\ \kappa\geqq1.
\end{cases}
\end{equation}
For simplicity, we shall absorb $\lambda$ into $C$ and $D$ and we write
\begin{equation}
\label{eq:excluded_volume_reduced}\lambda V_{exc}(\mathbf{\hat{l}}%
_{1},\mathbf{\hat{l}}_{2})=c-\frac{2d}{3}\left(  \mathbf{\bm{\sigma}(\hat{l}%
}_{1}\mathbf{)}:\mathbf{\bm{\sigma}(\hat{l}}_{2}\mathbf{)}\right)  ,
\end{equation}
where $c=\lambda C$ and $d=\lambda D$.

Substitution of the expression in Eq.~(\ref{eq:excluded_volume_reduced}) for
the excluded volume into Eq.~(\ref{eq_energy}) gives for the free energy
density (per unit volume)
\begin{equation}
\mathcal{F}=kT\rho_{0}\left\{  \ln\rho_{0}+\int_{\mathbb{S}^{2}}%
f(\mathbf{\hat{l}})\ln f(\mathbf{\hat{l}})d^{2}\mathbf{\hat{l}}-\int%
_{\mathbb{S}^{2}}f(\mathbf{\hat{l}})\ln\left[  1-\rho_{0}\left(  c-d\frac
{2}{3}\mathbf{\bm{\sigma}(\hat{l})}:\mathbf{Q}\right)  \right]  d^{2}%
\mathbf{\hat{l}}\right\}  ,
\end{equation}
where
\begin{equation}
\label{eq:order_tensor}\mathbf{Q}=\int_{\mathbb{S}^{2}}%
\mathbf{\bm{\sigma}(\hat{l})}f(\mathbf{\hat{l}})d^{2}\mathbf{\hat{l}%
}=\left\langle \bm{\sigma}(\mathbf{\hat{l}}) \right\rangle
\end{equation}
is the symmetric and traceless tensor orientational order parameter. The
eigenvectors of $\mathbf{Q}$ indicate the principal directions of average
orientation, and the eigenvalues provide a measure of the degree of order in
the direction along the corresponding eigenvector. The eigenvalues range from
$-\frac12$ to $1$.\footnote{In the mathematical literature, it is more
customary to define the order tensor $\mathbf{Q}$ as the ensemble average of
the orientation descriptor $\frac23\bm{\sigma}$. This makes the eigenvalues of
$\mathbf{Q}$ range from $-\frac13$ to $\frac23$, but it leaves unchanged the
definition (and the range) of the uniaxial scalar order parameter $S$ in
Eq.~\eqref{eq:uniaxial_scalar}.} Note that $f(\mathbf{\hat{l}})\geq0$ is
admissible if
\begin{equation}
1-\rho_{0}\left(  c-d\frac{2}{3}\mathbf{\bm{\sigma}(\hat{l})}:\mathbf{Q}%
\right)  >0,
\end{equation}
that is, if the argument of the logarithm is positive.

\subsection{Parametrization}

It is convenient to introduce the dimensionless parameter
\begin{equation}
\phi=\frac{\rho_{0}c-1}{\rho_{0}d}\in(-\infty,1], \label{eq_phi}%
\end{equation}
which is an increasing function of number density $\rho_{0}=\frac{1}{c-d\phi}%
$. In the very dilute limit, $\rho_{0}\rightarrow0^{+},\phi\rightarrow-\infty
$; in the dense packing limit, $\rho_{0}\rightarrow\rho_{0\max}=\frac{1}%
{c-d}^{-},\phi\rightarrow1^{-}$, this corresponds to the densest packing
fraction with only pairwise interactions. The free energy density can be
written in terms of $\phi$ as
\begin{equation}
\mathcal{F}=kT\rho_{0}\int_{\mathbb{S}^{2}}\left\{  f(\mathbf{\hat{l}})\ln
f(\mathbf{\hat{l}})-f(\mathbf{\hat{l}})\ln\left[  (1-\phi)-\left(  1-\frac
{2}{3}\mathbf{\bm{\sigma}(\hat{l}):Q}\right)  \right]  \right\}
d^{2}\mathbf{\hat{l}},%
\end{equation}
where we have neglected the additive constant $-kT\rho_{0}\ln d$. To
understand the physical significance of $\phi$, we note that we can write%
\begin{equation}
1-\phi=\frac{1}{\varphi}\left(  \frac{\rho_{0\max}}{\rho_{0}}-1\right),
\end{equation}
or%
\begin{equation}
1-\phi=\frac{1}{\varphi}\left(  \frac{v}{v_{\min}}-1\right),
\end{equation}
where $v=1/\rho_{0}$ is the volume per particle, and $v_{\min}=1/\rho_{0\max}$
is the minimum volume per particle. We regard the quantity $1-\phi$
as the the \emph{orientational }relative free volume, which provides a
dimensionless measure of the volume, or equivalently, of the number of states
available for orientation. If the particles are spheres, the anisotropy
$\varphi$ vanishes and the number of available states diverges. As the
anisotropy $\varphi$ is increased, the number of available orientational states
decreases, and the system is expected to become more and more ordered. Indeed,
this is what happens, as shown below.

\subsection{Minimization of the free energy}

We next minimize the free energy density with respect to the orientational
distribution function $f(\mathbf{\hat{l}})$. We write $\mathcal{F}$ explicitly
in terms of $f(\mathbf{\hat{l}})$,
\begin{equation}
\mathcal{F}=kT\rho_{0}\int_{\mathbb{S}^{2}}\left[  f(\mathbf{\hat{l}}_{1})\ln
f(\mathbf{\hat{l}}_{1})-f(\mathbf{\hat{l}}_{1})\ln\left(  \frac{2}%
{3}\mathbf{\bm{\sigma}(\hat{l}}_{1}\mathbf{)}:\int_{\mathbb{S}^{2}%
}\mathbf{\bm{\sigma} (\hat{l}}_{2}\mathbf{)}f(\mathbf{\hat{l}}_{2}%
)d^{2}\mathbf{\hat{l}}_{2}-\phi\right)  \right]  d^{2}\mathbf{\hat{l}}%
_{1}\mathbf{,}%
\end{equation}
where we have labelled the arguments for clarity.

We have two constraints: the distribution function must be normalized to
unity, that is,
\begin{equation}
\int_{\mathbb{S}^{2}}f(\mathbf{\hat{l}})d^{2}\mathbf{\hat{l}}=1,
\end{equation}
and the argument of the logarithm, the free volume fraction, must be positive;
that is
\begin{equation}
\label{eq:positivity_requirement}\frac{2}{3}\mathbf{\bm{\sigma}(\hat{l}%
)}:\mathbf{Q}-\phi>0.
\end{equation}

Setting formally the first variation to zero gives
\begin{equation}
\ln f(\mathbf{\hat{l}}_{1})\mathbf{+}(\mu+1) -\ln\left(  \frac{2}%
{3}\mathbf{\bm{\sigma}(\hat{l}}_{1}\mathbf{)}:\int_{\mathbb{S}^{2}%
}\mathbf{\bm{\sigma}(\hat{l}}_{2}\mathbf{)}f(\mathbf{\hat{l}}_{2}%
)d^{2}\mathbf{\hat{l}}_{2}-\phi\right)  -\bm{\sigma} \mathbf{\mathbf{(\hat{l}%
}}_{1}\mathbf{\mathbf{)}} :\left(  \int_{\mathbb{S}^{2}}f(\mathbf{\hat{l}}%
_{2})\frac{\frac{2}{3}\mathbf{\bm{\sigma} (\hat{l}}_{2})}{\frac{2}%
{3}\mathbf{\bm{\sigma}(\hat{l}}_{2}\mathbf{)}:\int_{\mathbb{S}^{2}%
}\mathbf{\bm{\sigma}(\hat{l}}_{3}\mathbf{)}f(\mathbf{\hat{l}}_{3}%
)d^{2}\mathbf{\hat{l}}_{3}-\phi}d^{2}\mathbf{\hat{l}}_{2}\right)  =0,
\end{equation}
where $\mu$ is a Lagrange multiplier associated with the normalization of
$f(\mathbf{\hat{l}})$. Solving for the distribution function gives the
self-consistent equation for $f$,
\begin{equation}
f(\mathbf{\hat{l}}_{1})=
\begin{cases}
\displaystyle \frac{\left[  \frac{2}{3}\mathbf{\bm{\sigma}(\hat{l}}%
_{1}\mathbf{)}:\int_{\mathbb{S}^{2}}\mathbf{\bm{\sigma}(\hat{l}}_{2}%
\mathbf{)}f(\mathbf{\hat{l}}_{2})d^{2}\mathbf{\hat{l}}_{2}-\phi\right]
e^{\bm{\sigma}\mathbf{\mathbf{(\hat{l}}}_{1}\mathbf{\mathbf{)}}:\int%
_{\mathbb{S}^{2}}f(\mathbf{\hat{l}}_{2})\frac{\frac{2}{3}%
\mathbf{\bm{\sigma}(\hat{l}}_{2})}{\frac{2}{3}\mathbf{\bm{\sigma}(\hat{l}}%
_{2}\mathbf{)}:\int_{S^{2}}\mathbf{\bm{\sigma}(\hat{l}}_{3}\mathbf{)}%
f(\mathbf{\hat{l}}_{3})d^{2}\mathbf{\hat{l}}_{3}-\phi}d^{2}\mathbf{\hat{l}%
}_{2}}}{\int_{\mathbb{S}^{2}}\Big\{\left[  (\frac{2}{3}%
\mathbf{\bm{\sigma}(\hat{l}}_{1}):\int_{\mathbb{S}^{2}}%
\mathbf{\bm{\sigma}(\hat{l}}_{2}\mathbf{)}f(\mathbf{\hat{l}}_{2}%
)d^{2}\mathbf{\hat{l}}_{2}-\phi)\right]  e^{\bm{\sigma} \mathbf{\mathbf{(\hat
{l}}}_{1}\mathbf{\mathbf{)}}:\int_{\mathbb{S}^{2}}f(\mathbf{\hat{l}}_{2}%
)\frac{\frac{2}{3}\mathbf{\bm{\sigma}(\hat{l}}_{2})}{\frac{2}{3}%
\mathbf{\bm{\sigma}(\hat{l}}_{2}\mathbf{)}:\int_{\mathbb{S}^{2}}%
\mathbf{\bm{\sigma}(\hat{l}}_{3}\mathbf{)}f(\mathbf{\hat{l}}_{3}%
)d^{2}\mathbf{\hat{l}}_{3}-\phi}d^{2}\mathbf{\hat{l}}_{2}}\Big\}d^{2}%
\mathbf{\hat{l}}_{1}}, & \text{if}\ \frac{2}{3}\mathbf{\bm{\sigma}(\hat{l}%
):Q}-\phi>0,\\
\  & \\
0, & \text{otherwise},
\end{cases}
\end{equation}
or in terms of the order parameter $\mathbf{Q}$,
\begin{equation}
f(\mathbf{\hat{l}}_{1})=
\begin{cases}
\displaystyle \frac{\left[  \frac{2}{3}\mathbf{\bm{\sigma}(\hat{l}}%
_{1}\mathbf{)}:\mathbf{Q}-\phi)\right]  e^{\bm{\sigma}\mathbf{\mathbf{(\hat
{l}}}_{1}\mathbf{\mathbf{)}}:\int_{\mathbb{S}^{2}}f(\mathbf{\hat{l}}_{2}%
)\frac{\frac{2}{3}\mathbf{\bm{\sigma} (\hat{l}}_{2})}{\frac{2}{3}%
\mathbf{\bm{\sigma}(\hat{l}}_{2}\mathbf{)}:\mathbf{Q}-\phi}d^{2}%
\mathbf{\hat{l}}_{2}}}{\displaystyle\int_{\mathbb{S}^{2}}\Big\{\left[  \frac{2}%
{3}\mathbf{\bm{\sigma}(\hat{l}}_{1}\mathbf{)}:\mathbf{Q}-\phi)\right]
e^{\bm{\sigma} \mathbf{\mathbf{(\hat{l}}}_{1}\mathbf{\mathbf{)}}%
:\int_{\mathbb{S}^{2}}f(\mathbf{\hat{l}}_{2})\frac{\frac{2}{3}%
\mathbf{\bm{\sigma}(\hat{l}}_{2})}{\frac{2}{3}\mathbf{\bm{\sigma}(\hat{l}}%
_{2}):\mathbf{Q}-\phi}d^{2}\mathbf{\hat{l}}_{2}} \Big\}d^{2}\mathbf{\hat{l}%
}_{1}}, & \text{if }\frac{2}{3}\mathbf{\bm{\sigma}(\hat{l}):Q}-\phi>0,\\
\  & \\
0, & \text{otherwise}.
\end{cases}
\end{equation}
It is convenient to define the tensor
\begin{equation}
\mathbf{\Psi=}\int_{\mathbb{S}^{2}}f(\mathbf{\hat{l}})\frac{\frac{2}%
{3}\mathbf{\bm{\sigma}(\hat{l}})}{\frac{2}{3}\mathbf{\bm{\sigma}(\hat{l}%
)}:\mathbf{Q}-\phi}d^{2}\mathbf{\hat{l}}= \left\langle \frac{\frac{2}%
{3}\mathbf{\bm{\sigma}}(\mathbf{\hat{l}} )}{\frac{2}{3}%
\mathbf{\bm{\sigma}(\hat{l})}:\mathbf{Q}-\phi}\right\rangle ,
\end{equation}
which can be regarded as an auxiliary order parameter. Finally, in terms of
both tensors $\mathbf{Q}$ and $\mathbf{\Psi}$, the expression for the
distribution function becomes
\begin{equation}
\label{eq:distribution_density}f(\mathbf{\hat{l}})=\frac{\big[\frac{2}%
{3}\mathbf{\bm{\sigma}(\hat{l}):Q}-\phi
\big]e^{\bm{\sigma}\mathbf{\mathbf{(\hat{l})}}:\mathbf{\Psi}}}{\int%
_{\mathbb{S}^{2}}\big[\frac{2}{3}\mathbf{\bm{\sigma}(\hat{l}):Q}%
-\phi\big]e^{\bm{\sigma} \mathbf{\mathbf{(\hat{l})}}:\mathbf{\Psi}}%
d^{2}\mathbf{\hat{l}}}.
\end{equation}
The requirement of positivity of the free volume fraction in
\eqref{eq:positivity_requirement} results in the orientational distribution
function being zero in some regions of orientation space. There are technical
issues surrounding the validity of using the Euler-Lagrange equation to find
minimisers of singular functionals like ours, but these have been addressed in
\cite{Jamie}, rigorously providing results fully consistent with ours.

\subsection{The equation of state}

To derive the equation of state of our system of hard ellipsoids, we start
with the relation between the pressure $P$ and the free energy density,
\begin{equation}
P=-\mathcal{F+}\rho_{0}\frac{\partial\mathcal{F}}{\partial\rho_{0}}.
\end{equation}
Writing out all the terms\footnote{It should be noted that the term omitted in
Eq.~\eqref{eq_14} would have brought an additional contribution $-kT\rho_{0}$
to the right side of Eq.~\eqref{eq:free_energy_density}, with no consequence
on the formula for $P$.}
\begin{equation}
\label{eq:free_energy_density}\mathcal{F}=kT\rho_{0}\left[  \ln\rho_{0}%
+\int_{\mathbb{S}^{2}}f(\mathbf{\hat{l}})\ln f(\mathbf{\hat{l}})d^{2}%
\mathbf{\hat{l}}-\int_{\mathbb{S}^{2}}f(\mathbf{\hat{l}})\ln\left(  1-\rho
_{0}c+\rho_{0}d\frac{2}{3}\mathbf{Q}:\mathbf{\bm{\sigma}(\hat{l})}\right)
d^{2}\mathbf{\hat{l}}\right]  ,
\end{equation}
and recalling Eq.~(\ref{eq_phi}) we get%
\begin{align}
\mathcal{F}  &  =\rho_{0}kT\left[  -\ln(d/2)+\int_{\mathbb{S}^{2}%
}f(\mathbf{\hat{l}})\ln f(\mathbf{\hat{l}})d^{2}\mathbf{\hat{l}}%
-\int_{\mathbb{S}^{2}}f(\mathbf{\hat{l}})\ln\left(  \frac{2}{3}\mathbf{Q}%
:\mathbf{\bm{\sigma}(\hat{l})}-\phi\right)  d^{2}\mathbf{\hat{l}}\right]
\mathbf{,}\\
\frac{\partial\mathcal{F}}{\partial\rho_{0}}  &  =\rho_{0}kT\left(
\int_{\mathbb{S}^{2}}\frac{f(\mathbf{\hat{l}})}{\frac{2}{3}\mathbf{Q}%
:\mathbf{\bm{\sigma}(\hat{l})}-\phi}d\mathbf{\hat{l}}\right)  \frac
{\partial\phi}{\partial\rho_{0}}\notag \\
&  +kT\left[  -\ln(d/2)+\int f(\mathbf{\hat{l}})\ln f(\mathbf{\hat{l}}%
)d^{2}\mathbf{\hat{l}}-\int f(\mathbf{\hat{l}})\ln\left(  \frac{2}%
{3}\mathbf{Q}:\mathbf{\bm{\sigma}(\hat{l})}-\phi\right)  d^{2}\mathbf{\hat{l}%
}\right]  ,\\
\frac{\partial\phi}{\partial\rho_{0}}  &  =\frac{1}{d\rho_{0}^{2}},\\
P  &  =\rho_{0}kT\left(  \frac{1}{\rho_{0}d}\int_{\mathbb{S}^{2}}%
\frac{f(\mathbf{\hat{l}})}{\frac{2}{3}\mathbf{Q:\bm{\sigma}(\hat{l})}-\phi
}d\mathbf{\hat{l}}\right)  =kT\frac{1}{d}\left\langle \frac{1}{\frac{2}%
{3}\mathbf{Q}:\mathbf{\bm{\sigma} (\hat{l})}-\phi}\right\rangle .
\end{align}
The latter can also be written as
\begin{equation}
P=\rho_{0}kT\left(  \frac{1}{1-\rho_{0}c}\right)  \left\langle \frac
{1}{1-\frac{2}{3}\phi^{-1}\mathbf{Q}:\mathbf{\bm{\sigma}(\hat{l})}%
}\right\rangle , \label{eq_pressure}%
\end{equation}
which shows the distinct, and at least formally equivalent, contributions of
positional and orientational entropy to the pressure.

If $\mathbf{Q}=0$, then
\begin{equation}
P=\frac{\rho_{0}kT}{1-\rho_{0}c},
\end{equation}
and if the ellipsoids are spheres, that is, if $\varphi=0$,
\begin{equation}
P=\frac{\rho_{0}kT}{1-\rho_{0}8\lambda v_{0}},
\end{equation}
which coincides with the van der Waals case without attractive interactions,
if we set $\lambda=\frac12$.

In general, Eq.~(\ref{eq_pressure}) is our equation of state. We shall show
numerically that in high density regime, when $\phi$ approaches $1$, the
pressure approaches infinity.

\section{The assumption of uniaxiality}

\label{sec:assumption_uniaxiality} Without external fields, classical models
(such as the Maier-Saupe model \cite{MaierSaupe58} for attractive interactions
and the Onsager model \cite{Onsager49} for steric interactions), predict only
uniaxial nematic order above the ordering transition \cite{Fatkullin05}. In
this section, we shall assume that $\mathbf{Q}$ is uniaxial for simplicity. We
demonstrate below that biaxial equilibrium also exists, but only as an
unstable saddle point, and therefore not physically observable. Under the
uniaxiality assumption, $\mathbf{Q}$ can be represented as
\begin{equation}
\mathbf{Q}=\frac{S}{2}(3\mathbf{\hat{n}\hat{n}-I}), \label{eq:uniaxial_scalar}%
\end{equation}
where $S$ is the scalar order parameter, providing a measure of the degree of
order, and $\mathbf{\hat{n}}$ is the nematic director, a unit vector
indicating the direction of average orientation. One can show that then
$\mathbf{\Psi}$ is also uniaxial, and shares the same eigenframe with
$\mathbf{Q}$, thus can be written as
\begin{equation}
\mathbf{\Psi}=\Psi(\mathbf{\hat{n}\hat{n}}-\frac{1}{3}\mathbf{I}).
\end{equation}
The uniaxiality assumption makes both the analysis and the numerics more
tractable. Now
\begin{equation}%
\begin{split}
\mathbf{Q}  &  :\mathbf{\bm{\sigma}}(\mathbf{\hat{l}})\mathbf{=}\frac{3S}%
{2}P_{2}(\mathbf{\hat{n}}\cdot\mathbf{\hat{l}}),\\
\mathbf{\Psi}  &  :\mathbf{\bm{\sigma}}(\mathbf{\hat{l}})=\Psi P_{2}%
(\mathbf{\hat{n}}\cdot\mathbf{\hat{l}}),
\end{split}
\label{eq:uniaxiality_assumption}%
\end{equation}
and, to within an inessential additive constant, the free energy density
becomes
\begin{align}
\mathcal{F}  &  =\rho_{0}kT\left(  \int_{\mathbb{S}^{2}}f(\mathbf{\hat{l}})\ln
f(\mathbf{\hat{l}})d^{2}\mathbf{\hat{l}}-\int_{\mathbb{S}^{2}}f(\mathbf{\hat
{l}})\ln[SP_{2}(\mathbf{\hat{n}}\cdot\mathbf{\hat{l}})-\phi]d^{2}%
\mathbf{\hat{l}}\right) \\
&  =\rho_{0}kT\left(  \Psi S-\ln\int_{\mathsf{S}_{+}}[SP_{2}(x)-\phi]e^{\Psi
P_{2}(x)}dx\right)  ,
\end{align}
where $x=\mathbf{\hat{n}\cdot\hat{l}}=\cos\theta$, $\mathsf{S}_{+}%
=\{x\in\lbrack-1,1]:SP_{2}(x)-\phi>0\}$, and use has been made of
Eq.~\eqref{eq:distribution_density} combined with
Eq.~\eqref{eq:uniaxiality_assumption}. The distribution function $f$, still
defined (and normalized) on $\mathbb{S}^{2}$, now only depends on $x$ and is
given by
\begin{equation}
f(x)=\frac{1}{2\pi}%
\begin{cases}
\displaystyle\frac{\left[  SP_{2}(x)-\phi\right]  e^{\Psi P_{2}(x)}}%
{\int_{\mathsf{S}_{+}}\left[  SP_{2}(x)\mathbf{-}\phi\right]  e^{\Psi
P_{2}(x)}dx}, & \text{if }SP_{2}(x)-\phi>0,\\
\  & \\
0, & \text{otherwise}.
\end{cases}
\label{eq:distribution_density_uniaxial}%
\end{equation}
Instead of solving the above self-consistent equation for $f(x)$, we solve the
coupled equations for $\Psi$ and $S$,
\begin{equation}
S=\frac{\int_{\mathsf{S}_{+}}P_{2}(x)\left[  SP_{2}(x)-\phi\right]  e^{\Psi
P_{2}(x)}dx}{\int_{\mathsf{S}_{+}}\left[  SP_{2}(x)\mathbf{-}\phi\right]
e^{\Psi P_{2}(x)}dx}=\left\langle P_{2}(x)\right\rangle , \label{eq_S_uni}%
\end{equation}%
\begin{equation}
\Psi=\frac{\int_{\mathsf{S}_{+}}P_{2}(x)e^{\Psi P_{2}(x)}dx}{\int%
_{\mathsf{S}_{+}}\left[  SP_{2}(x)-\phi\right]  e^{\Psi P_{2}(x)}%
dx}=\left\langle \frac{P_{2}(x)}{SP_{2}(x)-\phi}\right\rangle .
\label{eq_lambda_uni}%
\end{equation}
The set $\mathsf{S}_{+}$ is explicitly given by
\begin{align}
\mathsf{S}_{+}=(x_{0},1),\text{ if }S  &  >0,\label{eq:S_plus_first}\\
\mathsf{S}_{+}=(0,x_{0}),\text{ if }S  &  <0, \label{eq:S_plus_second}%
\end{align}
where $x_{0}$ is the \emph{positive}\footnote{We study $f$ only for $0\leq
x\leq1$, as by Eq.~\eqref{eq:distribution_density_uniaxial} it is even in
$-1\leq x\leq1$.} root of
\begin{equation}
SP_{2}(x_{0})-\phi=0. \label{eq:x_0_root}%
\end{equation}
To solve the above equations efficiently, by use of Eq.~\eqref{eq:x_0_root},
we eliminate $\phi$ from them, thus arriving at the following equations,
\begin{align}
S  &  =\frac{\int_{\mathsf{S}_{+}}P_{2}(x)\left[  P_{2}(x)-P_{2}%
(x_{0})\right]  e^{\Psi P_{2}(x)}dx}{\int_{\mathsf{S}_{+}}\left[
P_{2}(x)-P_{2}(x_{0})\right]  e^{\Psi P_{2}(x)}dx},\label{eq:S_equation}\\
\frac{1}{\Psi}  &  =\frac{\int_{\mathsf{S}_{+}}P_{2}^{2}(x)e^{\Psi P_{2}%
(x)}dx}{\int_{\mathsf{S}_{+}}P_{2}(x)e^{\Psi P_{2}(x)}dx}-P_{2}(x_{0}).
\label{eq:Psi_equation}%
\end{align}
The latter equation is solved for $x_{0}$, separately for the two cases in
Eqs.~\eqref{eq:S_plus_first} and \eqref{eq:S_plus_second}. For each of these
cases, we first assign a value for $\Psi\in(-\infty,\infty),$ then $x_{0}$ is
found by using the built in root finding routine in Wolfram Mathematica. Next,
the order parameter $S$ can be evaluated from Eq.~(\ref{eq:S_equation}).
Finally $\phi$ can be obtained from Eq.~\eqref{eq:x_0_root}. In this way, we
associate each value of $\Psi$ with the scalar order parameter $S$ and the
parameter $\phi$ expressing the excluded volume fraction.

\subsection{A special point}

The point $\Psi=0$ is a singular point of the above equations. In this case,
we consider Eq.~(\ref{eq_lambda_uni}) directly. It is easy to show that the
integration limit $x_{0}=0$ if $S>0$, whereas $x_{0}=1$ if $S<0$. By setting
$\Psi=0$ in Eq.~(\ref{eq_S_uni}), we get $\phi=-0.2$. To satisfy the
constraint
\begin{equation}
SP_{2}(x)-\phi>0,
\end{equation}
we immediately arrive at the following range for $S$,
\begin{equation}
-0.2<S<0.4.
\end{equation}
That is, for $\Psi=0$, corresponding to $\phi=-0.2$, any value of
$S\in(-0.2,0.4)$ is a solution of the Euler-Lagrange equation.

 \subsection{Homogeneous equilibrium solution}
\begin{figure}[th]
\centering
\includegraphics[width=.5\linewidth]{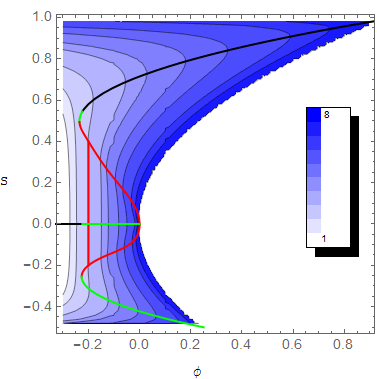} \caption{(color online)
Equilibrium order parameter $S$ vs.\ $\phi$ with uniaxial solutions only,
superimposed on the energy per particle contours. The black curve represents the stable
branch of the bifurcation, the green (light gray) curves correspond to the metastable
branches, and the red (dark gray) curves to the unstable branches.}
\label{fig-bifurcation}
\end{figure}

Fig.~\ref{fig-bifurcation} illustrates our key result. It shows the
equilibrium uniaxial order parameter $S$ vs.\ $\phi$. The stability of
different branches is determined by examining the local convexity of the free
energy density. To effectively illustrate this, the contours of free energy
per particle are superimposed on the bifurcation graph in the $S-\phi$ plane. The
black curve represents the stable branch. The green (light gray) curves correspond to
metastable states, and the red (dark gray) curves indicate unstable regimes. For
$\phi<\phi_{NI}=-0.224$, the system is in the isotropic state with $S=0$; at
$\phi=\phi_{NI},$ the system undergoes a first order transition to the nematic
state, with order parameter $S_{NI}=0.545$. As $\phi\rightarrow1$, the order
parameter $S\rightarrow1$ indicating a completely aligned configuration.
Inside the blue parabola are the regions where the order parameters of the
ordered state are not admissible. At $\phi=-0.2$, the vertical red (dark gray) line
indicates that all values of $-0.2<S<0.4$ share the same energy.

\begin{figure}[tbh]
\centering\includegraphics[width=.52\linewidth]{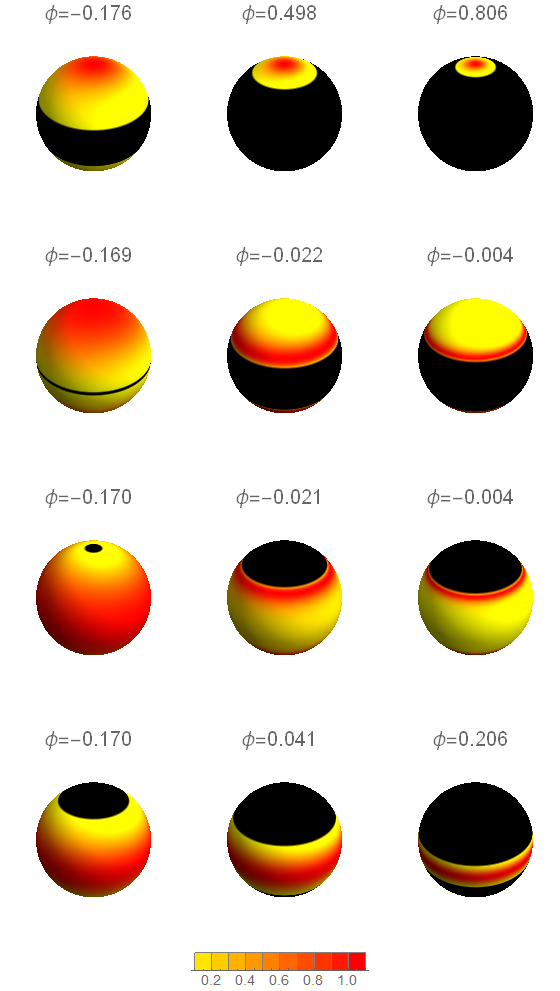} \caption{(color online) Density
plot of the orientational distribution function for selected equilibrium
uniaxial solutions. Top row: stable configuration with order parameter $S>0$;
Second row: unstable configuration with $S>0$; Third row: Unstable saddle
point configuration with $S < 0$; Bottom row: metastable configuration with
$S<0$.} \label{fig-pdf}
\end{figure}

\begin{figure}[th]
\centering
\subfigure[]{\includegraphics[width=0.48\linewidth]{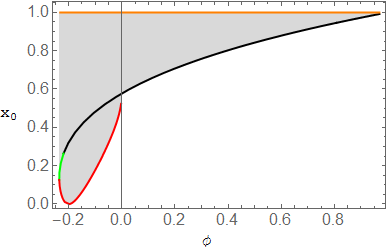}} \hspace
{.02\linewidth}
\subfigure[]{\includegraphics[width=0.48\linewidth]{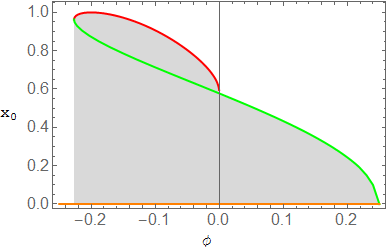}}
\caption{(color online)(a) Lower
integration limit $x_{0}$ vs.\ $\phi$ for uniaxial equilibrium solutions with
$S>0$; and the top horizontal line indicates the upper limit of the
integration. (b) Upper limit $x_{0}$ vs.\ $\phi$ for equilibrium solutions
with $S<0$; and the bottom horizontal line indicates the lower limit of the
integration. Black, green (light gray) and red (dark gray) colors correspond to stable, metastable, and
unstable solutions in Fig.~\ref{fig-bifurcation}.} \label{fig-x0}
\end{figure}

In Fig.~\ref{fig-pdf}, representative orientational probability density
functions are presented as a density plot on the surface of a unit sphere for
different equilibrium uniaxial solutions shown on Fig.~\ref{fig-bifurcation}.
In those graphs, the $z-$axis is chosen to be along the uniaxial director,
thus the density plots are axi-symmetric with respect to the $z-$axis. The
density functions are continuous, and normalized by their maximum values on
each sphere. The black regions on different spheres indicate forbidden
orientations, {\it i.e.}, no particles are allowed to orient in directions
corresponding to those regions. Yellow (light gray) indicates that only relatively few
particles are oriented in that direction, and red (dark gray) indicates that the majority
of particles are oriented in that direction. The figures clearly reveal that
the allowed regions of orientation shrink as the number density increases.
This effect is further illustrated in Fig.~\ref{fig-x0}.

Fig.~\ref{fig-x0} shows the lower (upper) limit $x_{0}$ of the integration for
$S>0$ $(S<0)$. The probability distribution function has no forbidden region
when $\Psi=0$. As one traces the stable black branch (Fig.~\ref{fig-x0}a) in the direction of
increasing $\phi$, the region shrinks. As $\phi\rightarrow1$, the lower limit
$x_{0}\rightarrow1$, which implies that all particles orient precisely in the
same direction, that is, they are perfectly ordered. At the lowest green (light gray)
branch ((Fig.~\ref{fig-x0}b)), as $\phi\rightarrow1/4$, the upper limit $x_{0}\rightarrow0,$ the
particles all lie in the plane perpendicular to the unique direction (normal
to the plane), but are randomly oriented in that plane. On the unstable red (dark gray)
branches (Fig.~\ref{fig-x0}ab), the particles are oriented inside or outside a cone, but more
towards the boundary of that cone, as shown in the second and third rows in
Fig.~\ref{fig-pdf}.

\subsection{Phase coexistence}

\label{phase_coexistence} For completeness, we next briefly inquire about the
possibility of coexisting nematic and isotropic phases in regions of $\phi$
where both isotropic and nematic solutions exist. We ask therefore whether the total free energy of
the system can be reduced by phase separation. Rather than plotting the free
energy density versus the number density, as is customary, In Fig.~4 we plot
the free energy per particle $\mathcal{F}/\rho_{0}$ versus $\phi$, and
implement the classical double tangent construction. Here the common slope
indicates equal pressures, and the linear dependence indicates equal chemical
potentials of the two coexisting phases. Using this representation, we learn
that the nematic and isotropic phases coexist with the universal dimensionless
parameter values $\phi_{I}=-0.3652$ and $\phi_{N}=0.1634$, regardless of
particle aspect ratio. The phase transition for the homogeneous nematic phase
occurs at $\phi_{NI}=-0.224$.
\begin{figure}[th]
\centering\includegraphics[width=.53\linewidth]{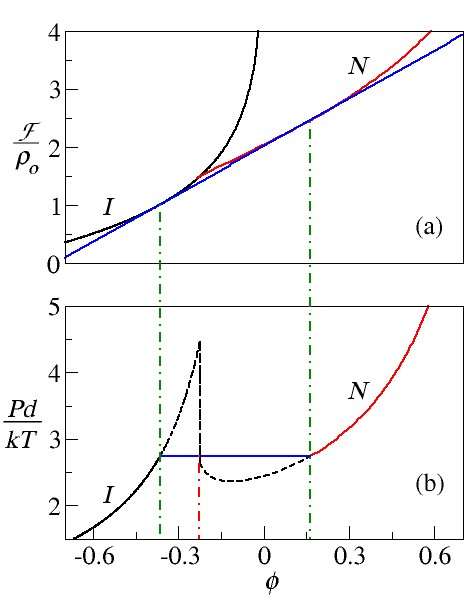}\caption{(color online) These diagrams illustrates two-phase coexistence: (a) shows the double tangent construction;
(b) shows the Maxwell, equal area construction. The dependence of pressure on $\phi$ in the isotropic and nematic phases is also shown in (b).}
\end{figure}

The volume fractions $\eta=\rho_{0}v_{0}$ of the two coexisting phases as
function of the aspect ratio $\kappa$ can be obtained at once from the values
of $\phi_{I}$, and $\phi_{N}$ though Eq.~\eqref{eq_phi}, once $\lambda$ is assigned.

\section{Biaxial equilibrium solutions}

\label{sec:biaxial} 
\begin{figure}[th]\centering
\includegraphics[width=.5\linewidth]
{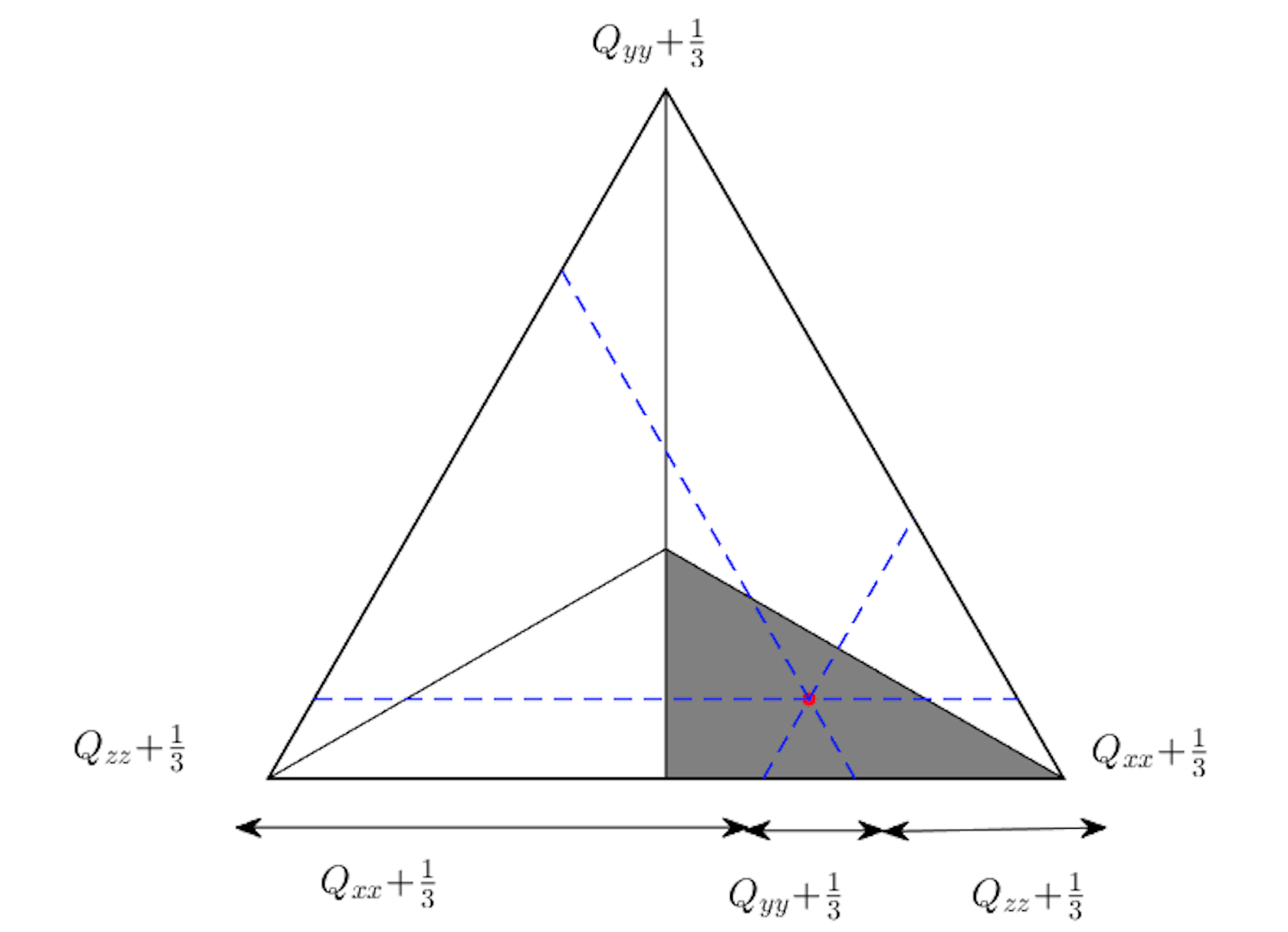} \caption{(color online) Ternary
diagram for representing $\mathbf{Q}$.}\label{fig-triangle}
\end{figure}

We anticipate the existence of biaxial equilibrium
solutions, especially for $\phi>0$, since the energy graph for each $\phi>0$
corresponds to two separated branches, with inadmissible regions in between.
The energy diverges as $S$ gets near the boundary of the inadmissible region
as seen in Fig.~\ref{fig-bifurcation}. If the system starts with a
configuration with $S<0$ near the metastable phase, it has to make its way to
the lowest energy state, and the only way is through biaxial phases. In its
eigenframe, $\mathbf{Q}$ can be represented as
\begin{equation}
\mathbf{Q}=\mathrm{diag}\left[  r\cos\left(  \alpha+\frac{2\pi}{3}\right)
,r\cos\left(  \alpha-\frac{2\pi}{3}\right)  ,r\cos\alpha\right]  ,
\end{equation}
thus $\mathbf{Q}$ is fully characterized by the parameters $r$ and $\alpha$
\cite{Zheng&P11}. We use a ternary diagram to represent $\mathbf{Q}$. Consider
an equilateral triangle with sides of unit length, centered on the origin,
with one vertex on the positive $y-$axis. Each point in the triangle
corresponds to a set of eigenvalues of $\mathbf{Q}$. To obtain the eigenvalues
of $\mathbf{Q}$, we draw lines parallel to each edge, through the point. Each
line intersects two edges, the length of the line segment from the point of
intersection to the vertex gives the eigenvalues of $\mathbf{Q}+\frac
13\mathbf{I}$. We illustrate this in Fig.~\ref{fig-triangle}. Due to symmetry,
we only consider the shaded $1/6^{th}$ portion of the triangle; and the rest
corresponds to $5$ different permutations of the same sets of eigenvalues of
$\mathbf{Q}$. Given two coordinates $(x,y)$, $0<x<\sqrt{3}/2$, $-1/2<y<\sqrt
{3}x/3$, of a point, $r$ and $\alpha$ can be obtained from
\begin{align}
r  &  =\sqrt{x^{2}+y^{2}}=\sqrt{\frac{2}{3}\mathbf{Q}:\mathbf{Q}},\\
\alpha &  =-\frac{\pi}{6}-\arctan\frac{y}{x}.
\end{align}
Here $r=0$, for which $\mathbf{Q}=\mathrm{diag}(0,0,0)$, corresponds to the
isotropic phase (the upper left corner of the shaded triangle in
Fig.~\ref{fig-triangle}); $\alpha=0,$ for which $\mathbf{Q}=\mathrm{diag}%
(-r/2,-r/2,r)$, corresponds to uniaxial prolate phases (the hypotenuse), and
$\alpha=\pi/3$ , for which $\mathbf{Q}=\mathrm{diag}(-r,r/2,r/2)$, corresponds
to uniaxial oblate phases (the short vertical edge); $\alpha=\pi/6,$ for which
$\mathbf{Q}=\mathrm{diag}(-\sqrt{3}r/2,0,\sqrt{3}r/2)$ , correspond to biaxial
phase with largest biaxiality.

\begin{figure}[th]
\centering
\subfigure[]{\includegraphics[width=.49\linewidth]{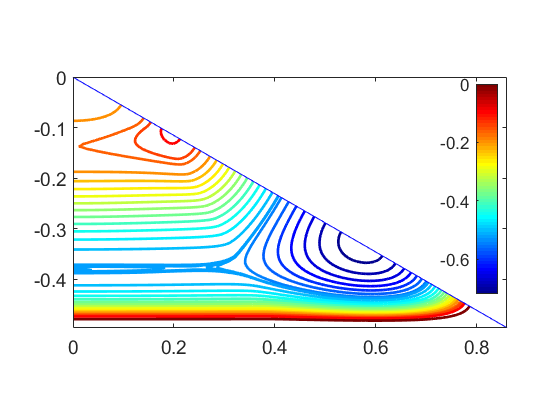}}
\hspace{.005\linewidth}
\subfigure[]{\includegraphics[width=.49\linewidth]{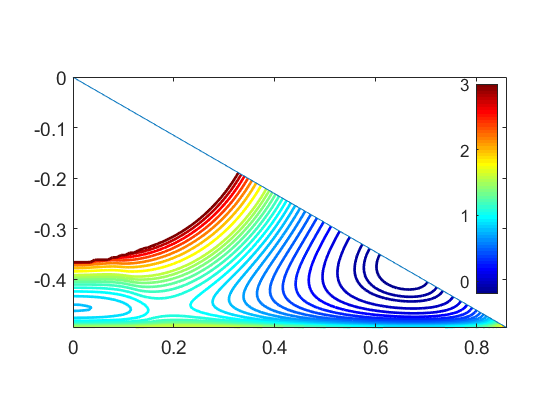}}
\caption{(color online) Energy
contours for two representative volume fractions. (a) $\phi=-0.1$; (b)
$\phi=0.1$.} \label{fig-biaxial}
\end{figure}

We select two representative values of $\phi$ to examine the landscape of the
free energy density vs.\ $\mathbf{Q}$. For each value of $\phi$, we specify
$\mathbf{Q}$, and numerically solve for $\mathbf{\Psi}$ using an iterative
Newton's method, and then numerically evaluate the energy density. In
Fig.~\ref{fig-biaxial}, energy density contours are plotted vs.\ different
$\mathbf{Q}$. For $\phi=-0.1$, there are six critical points on the energy
surface: one isotropic local minimizer, one prolate uniaxial global minimizer,
one uniaxial oblate local maximizer along the hypotenuse, one uniaxial oblate
local minimizer, one uniaxial oblate saddle point along the short vertical
edge, and one biaxial saddle point in the interior of the triangle. For
$\phi=0.1$, there are three critical points: one uniaxial prolate local
minimizer along the hypotenuse, one uniaxial oblate local minimizer along the
short vertical edge, and one biaxial saddle point in the interior of the
triangle. The energy blows up when $r^{2}<\phi$, so there is an inadmissible
region in the upper left portion of the triangle. In the regime where $\phi>0$, and the $S<0$ equilibrium solution is stable to biaxial perturbations, we expect a saddle-point corresponding to a ``mountain pass" between the $S>0$ and $S<0$ local minimizers. As we have demonstrated there are no other uniaxial critical points in this regime, this saddle-point must be biaxial. Such
behavior was not reported in Onsager's work, nor, as far as we know, in any
model for uniaxial particles without an external field.

\section{Comparison with simulations}

\label{sec:comparison} Since the early times of Frenkel and Mulder's
simulations with rotationally symmetric ellipsoids \cite{Frenkel85}, this
system of hard particles has become the test case for all theories that have
attempted to explain the behavior of dense nematics; we could not elude such a test.

Much is known now about the ordering transitions that take place in this
system as the number density is increased, including exotic crystal phases,
such as the newly discovered SM2 phase
\cite{donev:unusual,pfleiderer:simple,radu:solid}, which supplements the
stretched fcc-structure already known from \cite{Frenkel85}. New points on the
isotropic-to-nematic transition line of the phase diagram have also been added
for larger values of the aspect ratio $\kappa$ in \cite{camp:isotropic} and
\cite{odriozola:revisiting}. This latter also presents a full account on all
the phases presently known, enriching with new details the bell-shaped diagram
already hypothesized in \cite{Frenkel85} (see, in particular, Fig.~5 of
\cite{odriozola:revisiting}). Such a diagram has also proven to be relevant to
dynamical studies \cite{de_michele:dynamics,de_michele:simulation}, as its
shape is reflected by the rotational isodiffusivity lines in the $(\kappa
^{-1},\eta)$ plane, where $\eta=\rho_{0}v_{0}$ is the volume fraction. It was
shown in these studies how the isotropic-to-nematic transition is heralded by
a progressive hampering of the rotational dynamics, which would lead particles
to a complete arrest of the rotational motion, were this not pre-empted by the
ordering nematic transition.

Clearly, neither the crystal phases nor the dynamical precursors of the
nematic phase are within the reach of our theory. But we can still contrast
the latter with the simulation data available for the transition value
$\eta_{NI}$ of $\eta$.

\begin{figure}
[ptb]\centering
\includegraphics[width=.5\linewidth]{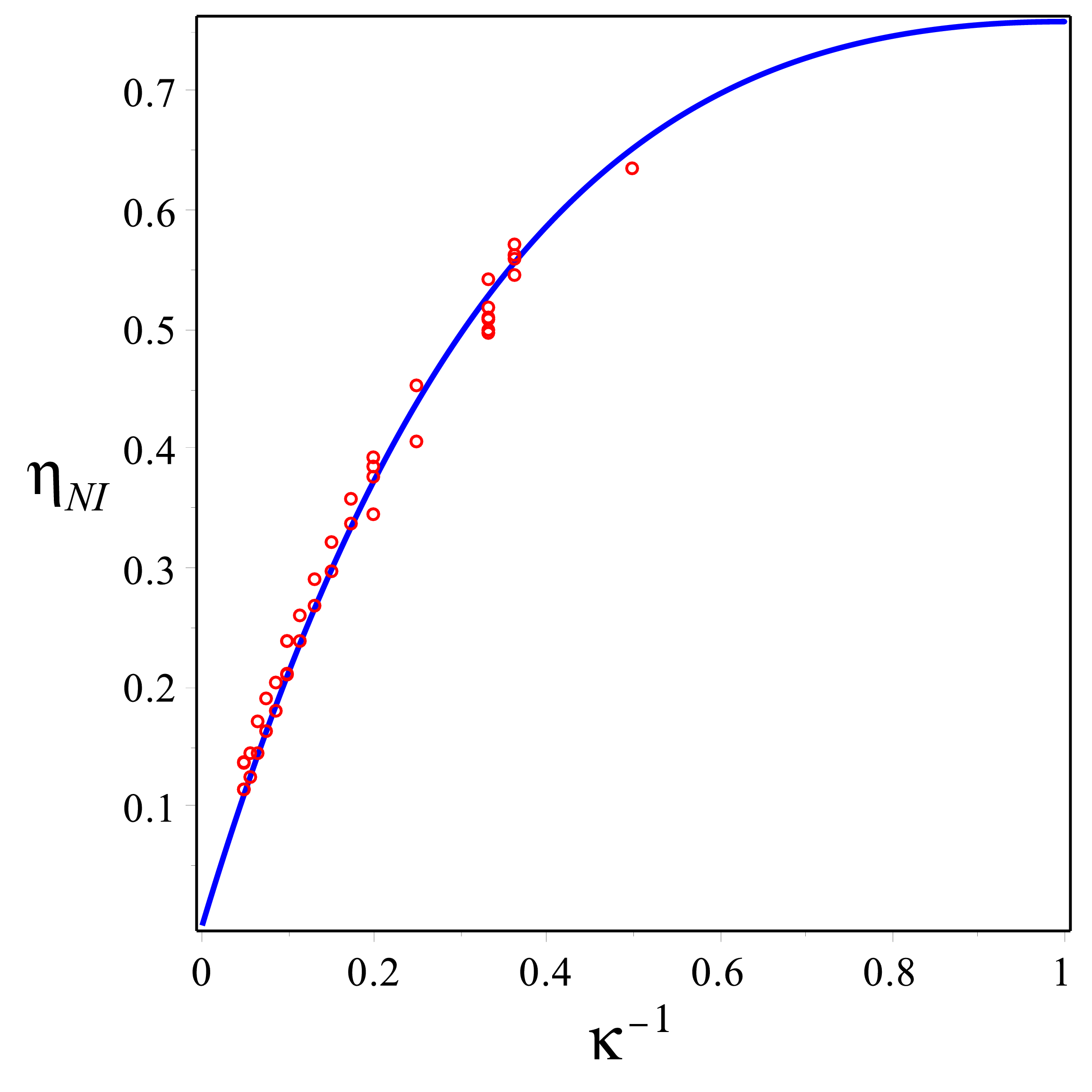}
\caption{(color online) The graph of $\eta_{NI}$ as in Eq.~\eqref{eq:eta_NI} against the reciprocal $\kappa^{-1}$ of the aspect ratio  for $\lambda$ as in Eq.~\eqref{eq:lambda_fit}. Simulation data are represented by red circles. Data have been collected from \cite{Frenkel85} and \cite{odriozola:revisiting} for both prolate and oblate ellipsoids, and reported here for the effective aspect ratio $\kappa>1$ (which is the real aspect ratio for prolate ellipsoids and its reciprocal for oblate ellipsoids). Data taken from \cite{Frenkel85} (see the first two columns of their Table 7) and from \cite{camp:isotropic} (see the fifth and eighth columns of their Table VI, where $\rho_\mathrm{cp}=\sqrt{2}$ ) include both ends of the coexistence range (and their $\eta$ values have been multiplied by $\pi/6$ to account for the fact that  the data  of \cite{Frenkel85} and \cite{camp:isotropic} are scaled to the volume $LW^2=6v_0/\pi$). Data taken from \cite{odriozola:revisiting} are all those in the first column of Table I, but the two referring to the cases 1.3:1-prolate and 1:1.3-oblate, as these latter fall inside the crystal region of the phase diagram (as also shown in Fig.~5 of \cite{odriozola:revisiting}).
} \label{fig:comparison}
\end{figure}

It readily follows from Eqs.~\eqref{eq_phi}, \eqref{eq_d}, and \eqref{eq_c}
that our theory predicts the following dependence for $\eta_{NI}$ on the
particles' anisotropy,
\begin{equation}
\eta_{NI}=\frac{1}{8\lambda}\frac{1}{1+(1-\phi_{NI})\varphi(\epsilon
)},\label{eq:eta_NI}%
\end{equation}
where $\varphi(\epsilon)$ is the shape function expressed explicitly by
Eq.~\eqref{eq:phi_function} in terms of the particles' eccentricity defined in
Eq.~\eqref{eq:eccentricity}. Letting $\phi_{NI}=-0.244$, as discussed above,
we can use Eq.~\eqref{eq:eta_NI} to determine $\lambda$ so as to fit the data
available from simulations for the isotropic-to-nematic transition. The
comparison of the graph of $\eta_{NI}$ with $41$ data points taken from
\cite{Frenkel85}, \cite{camp:isotropic}, and \cite{odriozola:revisiting} is
shown in Fig.~\ref{fig:comparison}.

Since $\eta_{NI}$ depends only on the eccentricity $\epsilon$ of the
ellipsoids, we have collected data for both prolate and oblate ellipsoids, and
plotted them for the effective aspect ratio $\kappa>1$ (which is the real
aspect ratio for prolate ellipsoids and its reciprocal for oblate ellipsoids).
The best least-squares fit is found for
\begin{equation}
\label{eq:lambda_fit}\frac1\lambda\doteq6.065,
\end{equation}
which is close to the value for close packed spheres in
Eq.~\eqref{eq:close_packed_spheres}. Setting $\frac{1}{\lambda}=6$ turns
Eq.~\eqref{eq:eta_NI} into our \emph{explicit}, theoretical prediction for the
isotropic-to-nematic transition line for a fluid of hard ellipsoids of revolution.

\section{Conclusions}

\label{sec:discussion} In this paper, we proposed a theory for dense nematics
whose constituting molecules interact only through the entropic forces arising
from mutual interpenetration. Our emphasis was on the behavior at high densities
and the effects of the depletion of available orientational states on
orientational order.

We studied a system of hard ellipsoidal particles and compared the predictions
of our theory with simulation data on the isotropic-to-nematic transition,
finding a good agreement for a specific value of a single fitting parameter
$\lambda$. This parameter should more generally be a function of the number density. Work to
determine the dependence of $\lambda$ on density is currently under way.

Our study was phrased in the canonical ensemble; we derived an expression for
the free energy at the van der Waals level. Low density expansions of the
logarithmic term agree with the free energy of Onsager \cite{Onsager49} and
Doi and Edwards \cite[p.~354]{Doi&Edwards}. A major challenge was the
determination of the orientational distribution function subject to the hard
constraint that the number states available to the system be positive. One
striking result of this constraint, in the mean field limit, is that at
densities above the isotropic-to-nematic transition, the orientational
distribution function which minimizes the free energy has a compact support
over orientation space. Although strictly forbidden orientations are likely an
artefact of our mean field approach, where particles effectively behave
identically, we expect that more sophisticated models would similarly show a
strong suppression of the corresponding orientational states.

We have found that the quantity $\phi$, related to the orientational relative
free volume, is a convenient control parameter characterizing the state of the
system, regardless of density or particle aspect ratio. Our model predicted a
first order nematic-isotropic phase transition; in a uniform system, the NI
transition occurs at $\phi_{NI}=-0.224$. In the regime $-0.3652\leq\phi$
$\leq0.1634$, the system is unstable, and undergoes phase separation. As
$\phi$ and the density increase, the regions of forbidden orientations grow,
and the degree of orientational order increases; the system becomes perfectly
aligned in the dense packing limit.

The equation of state from our free energy expression is more realistic than
Onsager's in that in our model, the pressure diverges in the dense packing
limit; our equation of state shows the individual contributions of positional
and orientational entropy to the pressure.

We see this work as providing a starting point towards a more systematic study
of dense nematic systems. Future work is aimed at identifying on a
first-principles basis the function of density that is to replace the single
parameter $\lambda$ that here ensured agreement between theory and simulation data.

\begin{acknowledgments}
X.Z. and P.P.-M. acknowledge support from NSF under DMS-1212046 and
EFRI-1332271. E.S.N. acknowledges support under FAPESP 2016/07448-5. J.M.T.
acknowleges support from European Research Council under the European Union's
Seventh Framework Programme (FP7/2007-2013)/ERC grant agreement $n^{o}$
291053. E.G.V. acknowledges the kind hospitality of the Oxford Centre for
Nonlinear PDE, where part of this work was done while he was visiting the
Mathematical Institute at the University of Oxford.
\end{acknowledgments}

\appendix

\section{Excluded volume of ellipsoids of revolution}

\label{app:excluded_volume} In this appendix we justify the approximate form
for the excluded volume of two congruent ellipsoids of revolution adopted in
this paper and we compare it with other approximations used in the literature.

Symmetry demands that the excluded volume of two congruent bodies of
revolution be a function $V_{exc}(x)$ of the inner product $x=\mathbf{\hat{l}%
}_{1}\cdot\mathbf{\hat{l}}_{2}=\cos\theta$ between the unit vectors
designating their axes. In general, $V_{exc}$ can be expanded in a series of
Legendre polynomials,
\begin{equation}
\label{eq:excluded_volume_expansion}V_{exc}(x)=\sum_{k=0}^{\infty}B_{k}%
P_{k}(x).
\end{equation}
It was shown in \cite{piastra:explicit} that for convex particles all $B_{k}$
can be expressed in terms of a countable family of extended invariant
Minkowski shape functionals, which can be computed \emph{explicitly} for
special shapes such as circular cones and ellipsoids of revolution. The
formula for $B_{0}$, which represents the isotropic average of $V_{exc}$, had
been known to be a function of the classical invariant Minkowski functionals
at least since the work of Isihara~\cite{isihara:determination}. It is easy to
show that all odd-indexed coefficients $B_{k}$ vanish identically for
particles that lack shape polarity.

For an ellipsoid of revolution with eccentricity $\epsilon$,
\begin{equation}
\label{eq:B_0}B_{0}=\frac32v_{0}\left[  \frac43 +\left(  1+(1-\epsilon
^{2})\frac{\arctanh\epsilon}{\epsilon} \right)  \left(  1+\frac{\arcsin
\epsilon}{\epsilon\sqrt{1-\epsilon^{2}}}\right)  \right]  ,
\end{equation}
\begin{equation}
\label{eq:B_2}B_{2}=\frac{15}{32}v_{0}\frac{1}{\epsilon^{4}}\left(
\epsilon^{2}-3+(\epsilon^{2}+3)(1-\epsilon^{2})\frac{\arctanh\epsilon
}{\epsilon}\right)  \left(  3-2\epsilon^{2}+\frac{4\epsilon^{2}-3}%
{\epsilon\sqrt{1-\epsilon^{2}}}\arcsin\epsilon\right)  ,
\end{equation}
where $v_{0}$ is the particle's volume. The former formula was already given
in \cite{isihara:determination}, while the latter formula coincides with that
given in \cite{isihara:theory} only for oblate ellipsoids (as pointed out in
\cite{piastra:explicit}, where Eq.~\eqref{eq:B_2} was established for all
ellipsoids of revolution, the formula of \cite{isihara:theory} for prolate
ellipsoids fails to be invariant under the transformation $\kappa
\mapsto1/\kappa$, and it is thus bound to be incorrect).

\begin{figure}[th]\centering
\subfigure[]{\includegraphics[width=0.32\linewidth]{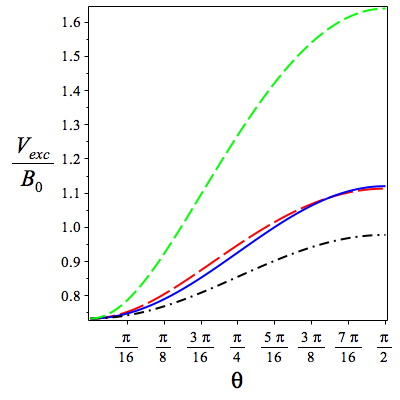}}
\subfigure[]{\includegraphics[width=0.32\linewidth]{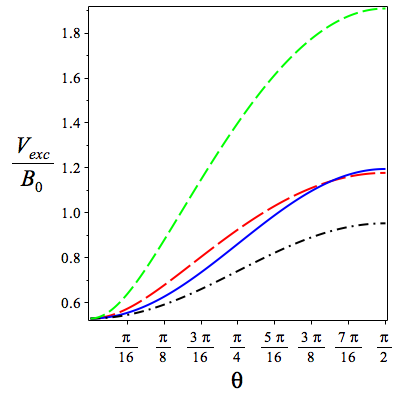}}
\subfigure[]{\includegraphics[width=0.32\linewidth]{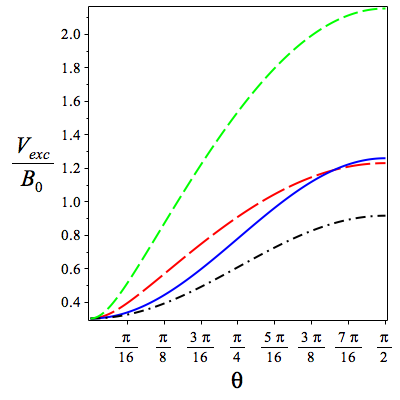}}
\caption{(color online) The excluded volume of a pair of congruent ellipsoids of revolution $V_{exc}$ (normalized to its isotropic average $B_0$) is plotted against the angle $\theta$ between their symmetry axes according to  four different approximate formulas: (red long-dashed) the expansion in Eq.~\eqref{eq:excluded_volume_expansion} truncated at $k=20$ with the coefficients computed according to the theory of \cite{piastra:explicit}; (blue solid) our approximate formula in Eq.~\eqref{eq:excluded_volume_app}; (black dash-dotted) Sheng's formula from \cite{Sheng75}; (green dashed) the HGO formula in Eq.~\eqref{eq:excluded_volume_HGO}. Values of the ellipsoids' aspect ratio: (a) $\kappa=3$, (b) $\kappa=5$, (c) $\kappa=10$.}
\label{fig:excluded_volume_comparison}
\end{figure}
We wish to approximate $V_{exc}(x)$ with
\begin{equation}\label{eq:excluded_volume_app}
\Vea(x)=C-DP_2(x),
\end{equation}
with $C$ and $D$ constants chosen in such a way that
\begin{equation}
\Vea(1)=8v_0
\end{equation}
and
\begin{equation}
\Vea(0)=B_0+B_2P_2(0)=B_0-\frac12B_2,
\end{equation}
where $B_0$ and $B_2$ are as in Eqs.~\eqref{eq:B_0} and \eqref{eq:B_2}. It is a simple matter to show that $C$ and $D$ can be written as in Eqs.~\eqref{eq_c} and \eqref{eq_d} with the shape function $\varphi$ defined as
\begin{equation}\label{eq:phi_function}
\begin{split}
\varphi(\ecc)=\frac{1}{768}\frac{1}{\ecc^6}\Big\{&254\ecc^6-135\ecc^4+135\ecc^2 +9\ecc(4\ecc^4+25\ecc^2-15)\frac{\arcsin\ecc}{\sqrt{1-\ecc^2}} \\
&+9\ecc(1-\ecc^2)(14\ecc^4+5\ecc^2-15)\arctanh\ecc\\
&+9(1-\ecc^2)(4\ecc^4-15\ecc^2+15)\frac{\arcsin\ecc}{\sqrt{1-\ecc^2}}\arctanh\ecc
\Big\}-\frac23.
\end{split}
\end{equation}
In the main text we have adopted the function in Eq.~\eqref{eq:excluded_volume_app} (and dropped the cumbersome superscript $^\mathrm{(app)}$). The function $\varphi$ is positive and monotonically increasing on the interval $[0,1)$; it diverges as $\ecc\to1^-$ and possesses the following asymptotic behaviors:
\begin{equation}
\lim_{\ecc\to0}\ecc^{-4}\varphi(\ecc)=\frac{1}{15},
\end{equation}
\begin{equation}
\lim_{\ecc\to1^-}\sqrt{1-\ecc^2}\varphi(\ecc)=\frac{21\pi}{256}.
\end{equation}

Other approximate formulae have been proposed in the past for the excluded volume of ellipsoids of revolution. Sheng~\cite{Sheng75} used a  formula like the one in Eq.~\eqref{eq:excluded_volume_app}, for which $\varphi$ would be replaced by the following simpler form,
\begin{equation}
\varphi(\ecc)=\frac{\left(\sqrt{1-\ecc^2}-1\right)^2}{\sqrt{1-\ecc^2}}.
\end{equation}
Berne and Penchukas~\cite{berne:gaussian} introduced the Hard Gaussian Overlap model (customarily abbreviated HGO in the literature), which mimics a short range repulsion between elongated molecules (see \cite{de_miguel:isotropic} for a lucid description of this model and its connections with traditional hard-particle models). When applied to hard ellipsoids, this model delivers an effective excluded volume written in the form
\begin{equation}\label{eq:excluded_volume_HGO}
V_{exc}^\mathrm{(HGO)}(x)=8v_0\sqrt{\frac{1-\chi^2x^2}{1-\chi^2}},\quad\text{where}\quad\chi^2=\frac{\kappa^2-1}{\kappa^2+1}.
\end{equation}

It is instructive to compare our approximate formula in Eq.~\eqref{eq:excluded_volume_app}, Sheng's variant, and the HGO formula with the remarkably good (at least for $\kappa\leq20$), but highly inconvenient expression obtained from Eq.~\eqref{eq:excluded_volume_expansion} by truncating the series expansion at $k=20$ and computing its coefficients \emph{exactly} through the theory presented in \cite{piastra:explicit}. Figure~\ref{fig:excluded_volume_comparison} illustrates such a comparison for $\kappa=3$, $5$, and $10$.
While the HGO formula highly overestimates the excluded volume, Sheng's formula underestimates it (though less dramatically so). By contrast, the formula we have used in this study seems to be more faithful, at least to  the truncated Legendre expansion.


\end{document}